\documentclass[prx,aps,showpacs,twocolumn,preprintnumbers,amsmath,amssymb,longbibliography,superscriptaddress]{revtex4-1}
\usepackage{amsmath,amssymb,amsfonts}
\usepackage{graphicx}
\usepackage[colorlinks=True,linkcolor=magenta,citecolor=blue,urlcolor=blue]{hyperref}
\usepackage{xcolor}
\usepackage{chngcntr}
\usepackage{braket}
\usepackage{hyperref}
\usepackage{nicefrac}
\usepackage{bm}
\usepackage{slashed}

\begin{document}
\bibliographystyle{apsrev4-1}
\title{Fermionic dualities with axial gauge fields}

\author{Adolfo G. Grushin}
\affiliation{Univ. Grenoble Alpes, CNRS, Grenoble INP, Institut N\'eel, 38000 Grenoble, France}
\author{Giandomenico Palumbo}
\affiliation{Center for Nonlinear Phenomena and Complex Systems,
Universit\'e Libre de Bruxelles, CP 231, Campus Plaine, B-1050 Brussels, Belgium}

\date{\today}
\begin{abstract}
The dualities that map hard-to-solve, interacting theories to free, non-interacting ones often trigger a deeper understanding of the systems to which they apply.
However, simplifying assumptions such as Lorentz invariance, low dimensionality, or the absence of axial gauge fields, limit their application to a broad class of systems, including topological semimetals.
Here we derive several axial field theory dualities in 2+1 and 3+1 dimensions by developing an axial slave-rotor approach capable of accounting for the axial anomaly. 
Our 2+1-dimensional duality suggests the existence of a dual, critical surface theory for strained three-dimensional non-symmorphic topological insulators.
Our 3+1-dimensional duality maps free Dirac fermions to Dirac fermions coupled to emergent U(1) and Kalb-Ramond vector and axial gauge fields.
Upon fixing an axial field configuration that breaks Lorentz invariance, this duality maps free to interacting Weyl semimetals, thereby suggesting that the quantization of the non-linear circular photogalvanic effect can be robust to certain interactions.
Our work emphasizes how axial and Lorentz-breaking dualities improve our understanding of topological matter.
\end{abstract}
\maketitle

\section{Introduction}

A defining property of massless relativistic fermions is that their momentum is either aligned or anti-aligned with their spin.
This quantum-mechanical degree of freedom is distinguished by axial gauge fields, which dramatically affect observables in a broad set of physical systems: from strained graphene and Weyl semimetals~\cite{amorim20161,Ilan:2019io}, to the quark-gluon plasma created in heavy ion collisions~\cite{Kharzeev2015}.
For example, in 3+1 dimensions the absence of axial charge conservation due to quantum fluctuations, known as the axial anomaly~\cite{B96}, significantly enhances the magnetoconductivity of Weyl semimetals~\cite{Armitage2018}.  
Within the quark-gluon plasma an axial chemical potential can generate a current parallel to a magnetic field, an otherwise absent phenomenon known as the chiral magnetic effect~\cite{Fukushima2008}.

Although axial gauge fields are physically ubiquitous, quantum field theory dualities are typically formulated without them. A quantum field theory duality is a map that renders two quantum field theories equivalent~\cite{Savit:1980kv}. They are especially useful when a strongly interacting theory that is hard to solve is mapped onto a free quantum field theory. An important recent example is the map proposed by Son~\cite{Son:2015gz} between a free 2+1-dimensional Dirac cone and 2+1-dimensional quantum electrodynamics (QED$_3$), see Ref.~\cite{Senthil:2019tl} for a review. It is a fermionic generalization of an older 2+1-dimensional boson-vortex duality~\cite{Peskin:1978jz,Halperin}, and its discovery suggested that the composite fermions describing the fractional quantum Hall state of a half-filled Landau level can be Dirac particles~\cite{Mross}. Son's fermionic duality has also been formulated as a duality between two surface theories, which correspond to two dual 3D topological insulator bulk theories~\cite{Metlitski,CWang}. This duality is embedded within a larger duality web~\cite{Seiberg:ur,Karch,Benini,Murugan:2017dr}, where different bosonic and fermionic theories can be related to each other by duality transformations.  There are variations that consider more than one fermionic flavor~\cite{Karch,Xu:2015ci,Sodemann17,Jensen:2019jv,Chen:2018cw,Potter2017}, as well as proposed extensions to 3+1 dimensions~\cite{Sagi:2018en,Palumbo2019,Bi,JWang,Furusawa:2019hx}. 

The description of a growing variety of systems in terms of axial gauge fields challenges us to develop dualities that can be used to understand their interacting phases. 
Moreover, it is known that the parity anomaly~\cite{Niemi:1983hw} is central to Son's 2+1-dimensional duality~\cite{Burkov2019}, yet a comparable understanding of the axial anomaly in putative 3+1 dimenisional fermionic dualities is still lacking. Our goal is to formulate dualities that help answer these questions.

\begin{figure}
    \centering
    \includegraphics[width=\columnwidth]{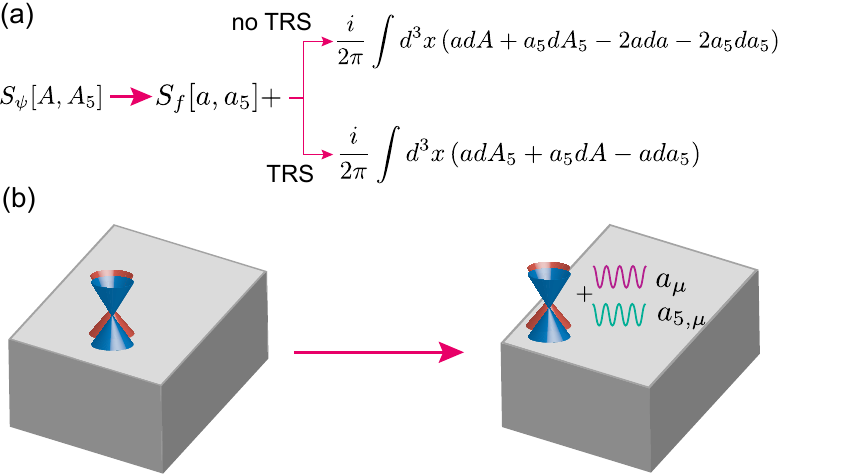}
    \caption{Schematic summary of the 2+1-dimensional axial dualities discussed in this work. (a) A theory of two helical massless fermions $\psi$ is coupled an external vector ($A_\mu$) and helical gauge fields ($A_{5,\mu}$). Physically, this can describe the double Dirac surface state of a strained non-symmorphic topological insulator. Depending on the realization of time-reversal symmetry, this theory maps to two different extensions of QED$_3$ of neutral fermions $f$, with mutual Chern-Simons coupling the two external fields with two emergent vector and helical fields, $a_\mu$ and $a_{5,\mu}$. (b) These dualities suggest the existence of dual surface theories for the double Dirac surface state of a strained non-symmorphic topological insulator.}
    \label{fig:2+1}
\end{figure}

\begin{figure}
    \centering
    \includegraphics[width=\columnwidth]{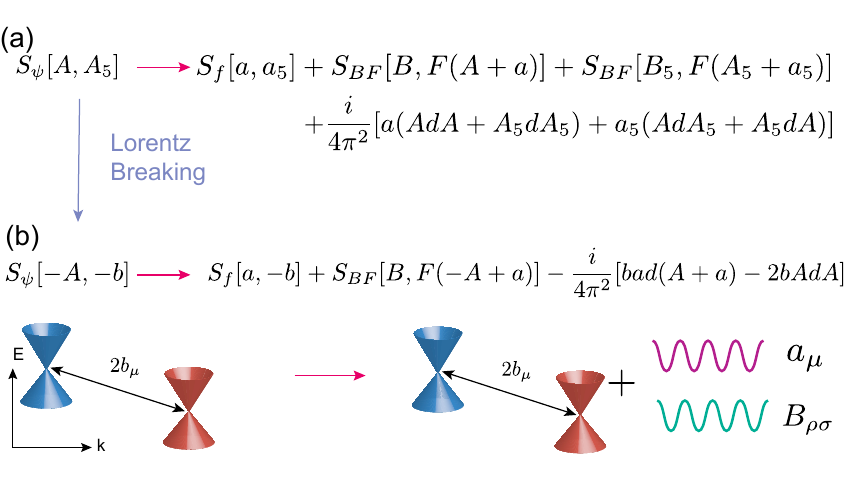}
    \caption{Schematic summary of the 3+1-dimensional axial dualities discussed in this work. (a) A massless Dirac fermion in 3+1-dimensions $\psi$, coupled to an external vector ($A_\mu$) and chiral ($A_{5,\mu}$) gauge fields is dual to a neutral Dirac fermion $f$ coupled to vector and axial dynamical gauge fields ($a_\mu$ and $a_{5,\mu}$), and a vector and axial Kalb-Ramond fields ($B_{\rho\sigma}$ and $B_{5,\rho\sigma}$) through terms of the form $\epsilon^{\mu\nu\rho\sigma}F_{\mu\nu}B_{\rho\sigma}$, known as BF terms. The last term accounts for the chiral anomaly. (b) When $A_{5,\mu}$ is set to a constant ($-b_\mu$), Lorentz symmetry is broken and the non-interacting theory describes a Weyl semimetal with Weyl node separation set by $b_\mu$. Its dual is an interacting Weyl semimetal theory with a BF, and mixed Carroll-Field-Jackiw terms. This duality suggests that an interacting Weyl theory can display a quantized photogalvanic effect.}
    \label{fig:3+1}
\end{figure}

In this work we derive several axial field theory dualities in 2+1 and 3+1 dimensions, summarized in Figs.~\ref{fig:2+1} and \ref{fig:3+1}. In 2+1 dimensions the helicity operator is well defined~\cite{Li}, unlike chirality~\footnote{ A note on wording: throughout our work we avoid the terminology of \textit{chiral gauge fields} in favour of \textit{axial gauge fields} to collectively refer to fields that couple with opposite signs to different helicities in 2+1 dimensions, or chiralities in 3+1 dimensions, since chirality is only well defined in even space-time dimensions.}. This implies that a (helical) gauge field can distinguish Dirac fermions by their helicity~\cite{Cortijo10}. The duality we derive maps two helical Dirac fermions coupled to external vector and helical gauge fields, into two helical Dirac fermions coupled to mixed Chern-Simons terms that couple the emergent vector and helical U(1) fields with the external fields. Our 2+1-dimensional duality suggests the existence of a surface theory dual to the surface Dirac fermion doublet found in strained 3D non-symmorphic topological insulators \cite{Wieder246}.

In 3+1 dimensions the duality we derive maps two Weyl fermions coupled to a vector ($A_{\mu}$) and an axial gauge field ($A_{5,\mu}$) to an interacting theory with two emergent U(1) vector fields ($a_\mu$ and $a_{5,\mu}$) and two emergent Kalb-Ramond fields ($B_{\mu\nu}$ and $B_{5,\mu\nu}$). The latter are anti-symmetric tensor gauge fields that originated in string theory \cite{Kalb,Banks}, and that appear in recent descriptions of 3+1-dimensional topological insulator theories~\cite{Cho,Fradkin,Cirio,Maciejko,Putrov}. Interestingly, our 3+1-dimensional duality applies to specific configurations of $A_{5,\mu}$ which describe different topological states, such as the 3D quantum Hall effect~\cite{Ramamurthy15,Galeski2020,thakurathi2020theory}, and Weyl semimetals \cite{Wan2011}. For example, the latter is recovered by choosing a constant $A_{5,\mu}$ on one side of the duality~\cite{Zyuzin2012,Grushin12,Zyuzin2012a,Goswami2013}, which breaks Lorentz symmetry and sets the Weyl node separation in momentum and energy space. In this case we find a duality between a Weyl semimetal, described by Lorentz-breaking QED with a constant axial four-vector~\cite{Colladay1997,Colladay1998,Grushin12}, and Lorentz breaking QED with a dynamical gauge field coupled to a Carroll-Field-Jackiw term \cite{Carroll90}. We show that this duality satisfies a requirement imposed by Son's fermionic duality.
The non-interacting side of our Weyl semimetal duality is known to display an exactly quantized circular photogalvanic effect~\cite{deJuan17}, a non-linear photocurrent generated by circularly polarized light. Our duality implies that the dual interacting theory must present the same quantized circular photogalvanic effect. This is in contrast to the effect of more conventional Coulomb interactions which correct the quantization constant if present~\cite{Avdoshkin20}.

To derive the dualities presented here we have developed an axial slave-rotor transformation that generalizes the slave-rotor technique~\cite{Florens:2002fz,Florens:2004er}, and incorporates the chiral anomaly in 3+1 dimensions. It is inspired by the work in Ref.~\cite{Burkov:2011de}, where this technique has been used to derive Son's duality and to emphasize the key role played by the parity anomaly \cite{Niemi:1983hw}. 
\section{Axial slave-rotor approach \label{sec:rotor}}
Our goal is to derive dualities between theories that contain two types of fermions, either with opposite helicity in 2+1 dimensions, or with opposite chirality in 3+1 dimensions. We therefore start by generalizing the slave-rotor approach~\cite{Florens:2002fz,Florens:2004er} to incorporate chirality and helicity. The method allows us to describe interactions in terms of two emergent Abelian gauge fields, and can be viewed as the U(1)$_V$ $\times$ U(1)$_A$ descendant of the SU(2) non-Abelian constructions in Refs.~\cite{Hermele:2007hb,Xu:2010ga}. In our case, U(1)$_V$ is associated to a vector gauge symmetry while U(1)$_A$ is associated to an axial gauge symmetry.

Our starting point is a system that can be decomposed into two independent sectors, that we call $L$ and $R$, such that the total Hilbert space $\mathcal{H}$ is
\begin{eqnarray}
\mathcal{H} \equiv \mathcal{H}_L \oplus \mathcal{H}_R.
\end{eqnarray}
Here $\mathcal{H}_{\chi}$ is the Hilbert space associated to each sector $\chi=L,R$.
These sectors are defined by the number operators at a given site $r$, $n_{\chi}$, which are independently conserved at classical level. To describe the physical fermions $\psi_{\chi}$, we introduce two independent rotor fields $\theta_{\chi}$, conjugate to $n_{r,\chi}$, satisfying the following relations

\begin{eqnarray}
\label{new}
\psi_{r,\chi} = e^{-i \theta_{r,\chi}} f_{r,\chi}, \hspace{0.3cm}\psi^{\dagger}_{r,\chi} = f^{\dagger}_{r,\chi}\, e^{i \theta_{r,\chi}}, 
\end{eqnarray}
such that
\begin{eqnarray}\label{conjugates2}
[\theta_{r,\chi}, n_{r,\chi}]=i.
\end{eqnarray}
The operators $ e^{i \theta_{r,\chi}}$ create a charged, spinless boson in the $\chi$ sectors. The operators $f^{\dagger}_{r,\chi}$ create neutral spinons that carry the electron's spin. In using Eq.~\eqref{new} we pay the price of enlarging the Hilbert space~\cite{Florens:2002fz,Florens:2004er}. To recover the physical Hilbert space in each sector it is necessary to impose the constraint
\begin{eqnarray}
\label{eq:constraint}
f^{\dagger}_{r,\chi}f_{r,\chi} =n_{r,\chi}+1.
\end{eqnarray}
These constraints act independently on each $\mathcal{H}_\chi$, and will be imposed at the level of the action with a Lagrange multiplier. As in previous works~\cite{Burkov2019,Palumbo2019}, we assume here 
that  $\langle e^{i \theta_{r,L}}\rangle \neq 0$ and $\langle e^{i \theta_{r,R}}\rangle \neq 0$, which implies the absence of a Mott insulating phase~\cite{Pesin:2010va}.

\section{2+1 Fermion-Fermion duality with an axial gauge field \label{sec:2+1}}

\subsection{Formulation of the duality\label{sec:form2+1}}

In 2+1 dimensions there is no notion of chirality. However, two Dirac fermions can form a reducible $4\times 4$ representation of the Clifford algebra, such that the $2\times 2$ irreducible blocks that compose this representation can be labeled by their helicity (left and right), to which an axial gauge field can couple to. In what follows we use the axial slave-rotor approach presented in the previous section to connect two theories involving massless Dirac fermions of opposite helicities. The first theory is a non-interacting theory of two helical massless Dirac fermions in Euclidean spacetime, defined as
\begin{subequations}
\label{eq:Diracall}
\begin{eqnarray}
\label{eq:Dirac}
S_{h}&=&\int d^3 x\, \bar{\psi} \gamma^\mu ( \partial_{\mu}-i A_{\mu}-i A_{5,\mu}\gamma_5) \psi\\
\nonumber
&\equiv&\int d^3 x\, \left[\bar{\psi}_L\, \sigma^\mu_{L} ( \partial_{\mu}-i A_{\mu,L}) \psi_L\right.\\
&+&\left.\bar{\psi}_R\, \sigma^\mu_R ( \partial_{\mu}-i A_{\mu,R}) \psi_R\right],
\end{eqnarray}
\end{subequations}
where $A_{\mu}$ is an external electromagnetic field, $A_{5,\mu}$ is an axial gauge field, $\psi=(\psi_L, \psi_R)^T$ is a four-component spinor, $\sigma^\mu_L=(\mathbb{I}, \sigma^i)$, ${\sigma}^\mu_R=(\mathbb{I}, -\sigma^i)$, and
\begin{subequations}
\begin{eqnarray}
\label{eq:fields}
A_{\mu,L}=A_{\mu}+A_{5,\mu},\\
A_{\mu,R}=A_{\mu}-A_{5,\mu}.
\end{eqnarray}
\end{subequations}
We find this theory to be dual to neutral Dirac fermions $f$ coupled to an emergent vector and axial gauge field, $a_\mu$ and $a_{5,\mu}$, respectively. These emergent gauge fields are coupled with the external $A_{\mu}$ and $A_{5,\mu}$ fields through mixed Chern-Simons terms. If time-reversal symmetry is absent, then the dual theory to Eqs.~\eqref{eq:Diracall} takes the following form
\begin{subequations}
\label{eq:cQED3i}
\begin{eqnarray}
\label{eq:cQED3ai}
\nonumber
S^{(1)}_{\mathrm{cQED_{3}}}&=&\int d^3 x\, \bar{f} \gamma^\mu ( \partial_{\mu}-i a_{\mu}-i\gamma_5 a_{5,\mu}) f \\
&+& \dfrac{i}{2\pi}adA + \dfrac{i}{2\pi}a_{5}d A_{5}-\dfrac{i}{4\pi}ada - \dfrac{i}{4\pi}a_{5}d a_{5}+\cdots.\nonumber \\
\end{eqnarray}
Here we make use of the short hand differential form notation $ada = \epsilon^{\mu\nu\rho}a_\mu\partial_\nu a_\rho$. 
In the ellipses ($+\cdots$) we include higher-derivative kinematic Maxwell terms, which can be neglected to lowest order, and are not relevant for our discussion.  When the axial gauge fields are switched off, the duality between Eqs.~\eqref{eq:Diracall} and \eqref{eq:cQED3ai} reduce to two copies of Son's duality \cite{Son:2015gz,Sodemann17}, one for each helicity. 
If time-reversal symmetry is preserved then the dual theory of Eqs.~\eqref{eq:Diracall} is given by
\begin{eqnarray}
\label{eq:cQED3bi}
\nonumber
S^{(2)}_{\mathrm{cQED_{3}}}&=&\int d^3 x\, \bar{f} \gamma^\mu ( \partial_{\mu}-i a_{\mu}-i\gamma_5 a_{5,\mu}) f \\
&+&  \dfrac{i}{2\pi}adA_5 + \dfrac{i}{2\pi}a_{5}d A- \dfrac{i}{2\pi}ad a_{5}+\cdots.
\end{eqnarray}
These are the main results of this section, and are summarized in Fig.~\ref{fig:2+1}.

\end{subequations}

\subsection{Derivation of the duality}

We begin by defining a lattice version of the gapless Hamiltonian of Eq.~\eqref{eq:Diracall}, given by two decoupled Hamiltonians
\begin{eqnarray}
H=H_{L}+H_{R},
\end{eqnarray}
where the Hamiltonian of each sector is given by
\begin{eqnarray}\label{Dirac}
\nonumber
H_{\chi}&=& \sum_{r}\left[\psi^{\dagger}_{r,\chi}\left(\frac{-i\hat{\chi}\sigma^{s}-m_{\chi} \sigma^{z} }{2}\right)e^{-i A_{r,r+\hat{s},\chi}}\psi_{r+\hat{s},\chi}+{\rm h.c.}\right] \\ 
&+&\sum_{r}\psi^{\dagger}_{r,\chi}\left[(m_{0}+2m_{\chi})\sigma^z - i A_{0,r,\chi}\right]\psi_{r,\chi}.
\end{eqnarray}
Here $r=\{x,y\}$ is the site index, $A_{r,r+\hat{s},\chi}$ is introduced through a Peierls substitution on the lattice link $(r, r+\hat{s})$ with $\hat{s}\equiv (\hat{x},\hat{y})$. We have also introduced the scalar $\hat{\chi}$ that takes the value $\hat{\chi}=+1$ and $\hat{\chi}=-1$ for chiralities $\chi=L$  and $\chi=R$, respectively. The parameter $m_0$ sets the gap at the $\Gamma$ point, while a combination of $m_0$ and $m_{\chi}$ sets the gaps at momenta ${\bf{K}}_{i}=(0,\pi),(\pi,0)$ and $(\pi,\pi)$. When $m_0=0$, the low-energy theory around $\Gamma$ takes the form of a gapless Dirac fermion %
\label{eq:lowen}
\begin{eqnarray}
H_{\chi}&=& \hat{\chi}\int d^{3}k  \hspace{1mm} \bar{\psi}_{k,\chi}\hspace{1mm} \sigma^{i}k_{i} \hspace{1mm}\psi_{k,\chi}.
\end{eqnarray}
For a finite $m_0$, the theory becomes that of a massive Dirac fermion that may be integrated out. The resulting effective field theory takes the form of a Chern-Simons theory~\cite{Niemi:1983hw}
\begin{eqnarray}
\label{eq:CSburkovs0}
\nonumber
    S_\mathrm{eff}&=& -i\dfrac{\mathrm{sgn}(m_{0})}{8\pi}\int d^3 x\;  \epsilon^{\mu\nu\rho}A_{\mu,\chi}\partial_{\nu}A_{\rho,\chi} \\
    &\equiv&-i\dfrac{\mathrm{sgn}(m_{0})}{8\pi} \int d^3 x\; A_{\chi}dA_{\chi}.
\end{eqnarray}
where in the second line we have defined a short-hand notation for the Chern-Simons term.

Accordingly, the low-energy theories around ${\bf{K}}_{i}$ are gapped Dirac fermions with masses set by a combination of $m_0$ and $m_{\chi}$. We are interested in the limit of small $m_0$, for which the combined effective action including ${\bf{K}}_{i}$ is~\cite{Niemi:1983hw}
\begin{eqnarray}
\label{eq:CSburkovs}
    S_\mathrm{eff,\chi}&=& i\dfrac{\mathrm{sgn}(m_{\chi})}{8\pi} \int d^3x\; A_{\chi}dA_{\chi}.
\end{eqnarray}
The role of time-reversal symmetry is explicit when we set $A_{5,\mu}=0$, and therefore $A_{\mu,L}=A_{\mu,R}=A_\mu$.
If a Chern-Simons term $AdA$ is present in the total effective action, this implies a finite Hall conductivity and the breaking of time-reversal symmetry. 
The total effective action is obtained by combining Eq.~\eqref{eq:CSburkovs0} and Eq.~\eqref{eq:CSburkovs} for both $\chi$ sectors:
\begin{eqnarray}\label{CSeff}
\nonumber
S_{\mathrm{eff}}&=&\frac{i}{8 \pi} \int d^3 x\, \left(-\mathrm{sgn}(m_0)+\mathrm{sgn}(m_{L})\right) AdA\\
&\pm&\left(-\mathrm{sgn}(m_0)+\mathrm{sgn}(m_{R})\right) AdA.
\end{eqnarray}
The relative sign between $m_0$ and $m_{\chi}$ determines whether a Chern-Simons term $AdA$ is allowed within each sector, and therefore sets their respective Hall conductivities. The $\pm$ represents the freedom to choose the relative sign between the Hall conductivities of the $L$ and $R$ sectors.

Since each $\chi$ sector is described by the Hamiltonian considered in Ref. \cite{Burkov2019}, we can use the two independent slave-rotor transformations, introduced in the previous section, to find the dual theory of Eqs.~\eqref{eq:Diracall}. For each $\chi$, the derivation follows the method in Ref.~\cite{Burkov2019}, but here we will keep track of the signs of the different Chern-Simons terms that are induced via the parity anomaly. Because both sectors remain decoupled, we detail the derivation for the dual action for the $L$ sector only. At the end, we will combine both chiral sectors into a single theory by considering the role of time-reversal symmetry. 

Using the slave-rotor transformation Eq.~\eqref{new} we can write the imaginary-time action ($\tau=i t$) that corresponds to $H_{L}$ as
\begin{eqnarray}\label{action1s}
S=\int_{0}^{\beta}d\tau \sum_{r,s}\left[f_{r,L}^{\dagger}\partial_{\tau} f_{r,L}
-i n_{r,L}(\partial_{\tau}\theta_{r,L}+A_{r,0,L}) \right. \hspace{0.4cm} \nonumber \\ 
\left. +(m_{0}+2m_{L}) f_{r,L}^{\dagger} \sigma_z f_{r,L}  + i \lambda_{r,L}(f_{r,L}^{\dagger}f_{r,L}-n_{r,L}-1) \right. \nonumber \\ \left.
+f_{r,L}^{\dagger}\left(\frac{-i\sigma^{s}-m_L \sigma^{z} }{2}\right)e^{-i( A_{r,r+\hat{s},L}+\Delta_s \theta_{r,L})}f_{r+\hat{s},L}
\right. \nonumber \\ \left. 
+{\rm h.c.}\right], \hspace{0.5cm}
\end{eqnarray}
where $\Delta_s \theta_{r,L} =  \theta_{r+\hat{s},L} - \theta_{r,L}$ and $\lambda_{r,L}$ is a Lagrange multiplier field that imposes the constraint Eq.~\eqref{eq:constraint}. 
To decouple the $f$ fermions from the rotor and external gauge fields, $\theta_{r,\chi}$ and $A_{r,\chi}$, respectively, we introduce a Hubbard-Stratonovich field $h_L\equiv\zeta_L e^{i a_L}$ defined on the lattice~\cite{Lee:2005ew,Barkeshli:2012fx}. 
Since the amplitude fluctuations are gapped, we can fix the magnitude $\zeta_{L}$ to its saddle point value and consider only phase fluctuations. In this case, Eq.~(\ref{action1s}) can be rewritten as follows:
\begin{widetext}
\begin{eqnarray}\label{action2s}
S&=&\int_{0}^{\beta}d\tau \sum_{r,s}\left[f_{r,L}^{\dagger}(\partial_{\tau}+i a_{r,0,L}) f_{r,L} -i n_{r,L}(\partial_{\tau}\theta_{r,L}+A_{r,0,L}+a_{r,0,L})+(m_{0}+2m_{L})f_{r,L}^{\dagger} \sigma_z f_{r,L}
 \right. \nonumber \\ 
&+&\left.\left[\zeta_L f_{r,L}^{\dagger}\left(\frac{-i\sigma^{s}-m_L \sigma^{z} }{2}\right)e^{i a_{r,r+\hat{s},L}}f_{r+\hat{s},L}+{\rm h.c.}\right]
-\zeta_L \cos \left(\Delta_{\hat{s}}\theta_{r,L}+A_{r,r+\hat{s},L}+a_{r,r+\hat{s},L}\right)
\right],
\end{eqnarray}
\end{widetext}
where we have identified the Lagrange multiplier with the temporal component of the emergent gauge field, $a_{r,0,L}\equiv \lambda_{r,L}$. We now may use the Villain approximation to approximate the last cosine as~\cite{Villain75}
\begin{equation}
    e^{\zeta \cos(\alpha)} \approx \sum_{J} e^{-iJ \alpha-(1/2\zeta)J^2},
\end{equation}
at the expense of introducing a boson current $J^{r,r+\hat{s}}_L$. After this step, the full action consists of two terms
\begin{eqnarray}
S=S_{f}+S_{\theta},
\end{eqnarray}
where
\begin{eqnarray}\label{action3s}
\nonumber
S_f&=&\int_{0}^{\beta}d\tau \sum_{r,s}\left[f_{r,L}^{\dagger}(\partial_{\tau}+i a_{r,0,L}) f_{r,L} \right.\\
\nonumber
&+& \left. (m_{0}+2m_{L}) f_{r,L}^{\dagger} \sigma_z f_{r,L} \right. \nonumber \\
&+&\left.  \zeta_L f_{r,L}^{\dagger}\left(\frac{-i\sigma^{s}-m_L \sigma^{z}}{2}\right)e^{i a_{r,r+\hat{s},L}}f_{r+\hat{s},L}+{\rm h.c.}\right],\nonumber \\
\end{eqnarray}
and
\begin{eqnarray}
\label{S2s}
S_{\theta}&=&
\int_{0}^{\beta}d\tau \sum_{r,s}\left[i J^{r,0}_L(\partial_{\tau}\theta_{r,L}+A_{r,0,L}+a_{r,0,L})   \right.\nonumber  \\ 
\nonumber
&+& i J^{r,r+\hat{s}}_L \left(\Delta_{\hat{s}}\theta_{r,L}+A_{r,r+\hat{s},L}+a_{r,r+\hat{s},L}\right)\\ &+&\left. \dfrac{1}{2\zeta_L}\left(J^{r,r+\hat{s}}_L\right)^{2} \right],
\end{eqnarray}
where we have identified $n_{r,L}$ as the temporal component ($J^{r,0}_L$) of the bosonic current $J^{r,r+\hat{s}}_L$.
In the continuum limit (i.e., long-wavelength limit) when $m_0=0$, Eq.~(\ref{action3s}) becomes
\begin{eqnarray}
S_{f}=\int  d^{3}x\, \bar{f}_L\, \sigma^{\mu}_L(\partial_{\mu}+ia_{\mu,L})f_L.
\end{eqnarray}
In this limit, $\Delta_{\hat{s}}$ becomes the standard spatial derivative and by integrating out $\theta_L$ in Eq. (\ref{S2s}), we obtain
\begin{eqnarray}
\partial_{\mu}J^{\mu}_L=0.
\end{eqnarray}
A solution to this equation is given by
\begin{eqnarray}
J^{\mu}_L=\frac{1}{4 \pi}\epsilon^{\mu\lambda\nu}\partial_{\nu}b_{\lambda,L},
\end{eqnarray}
which we can insert back into Eq.~\eqref{S2s}. The low-energy action now reads
\begin{eqnarray}
\nonumber
S&=&\int d^3 x\, \left[\bar{f}_L\, \sigma^\mu_L ( \partial_{\mu}+i a_{\mu,L}) f_L\right]-\frac{i}{4 \pi} b_{\mu,L}d(a_L+A_L)\\
&+&\frac{1}{128 \pi^2 \zeta_L} F(b_L)_{\mu\nu}F(b_L)^{\mu\nu},
\end{eqnarray}
where we have defined the field-strength as $F(b)_{\mu\nu}=\partial_{\mu}b_{\nu}-\partial_{\nu}b_{\mu}$.

To obtain an action with a single statistical field, we first separate the fermionic high- and low-energy modes. The former are gapped and can be integrated out by the help of Eq.~\eqref{eq:CSburkovs} resulting in
\begin{eqnarray}
\nonumber
S&=&\int d^3 x\, \left[\bar{f}_L\, \sigma^\mu_L ( \partial_{\mu}+i a_{\mu,L}) f_L\right]-\frac{i}{4 \pi} b_Ld(a_L+A_L)\\
\label{eq:S3s}
&+&i\frac{\mathrm{sgn}(m_L)}{8 \pi} a_Lda_L+\cdots.
\end{eqnarray}
The ellipses in last line contain the kinematical Maxwell term, which is of higher-order in derivatives and can be neglected in the low-energy limit. We can now integrate out one of the statistical gauge fields, keeping track of the mass (see Appendix.~\ref{app:intout}). This amounts to the replacement
\begin{eqnarray}
b_{\mu,L}\rightarrow \mathrm{sgn}(m_L)a_{\mu,L}, 
\end{eqnarray}
which delivers
\begin{eqnarray}
\nonumber
S&=&\int d^3 x\, \left[\bar{f}_L\, \sigma^\mu_L ( \partial_{\mu}+i a_{\mu,L}) f_L\right]-i\frac{\mathrm{sgn}(m_L)}{4 \pi} a_Ld A_L \\
\label{eq:S3s2}
&-&i\frac{\mathrm{sgn}(m_L)}{8 \pi} a_Lda_L+\cdots.
\end{eqnarray}
Finally, by combining the $L$ and $R$ sectors, we arrive at
\begin{eqnarray}
\nonumber
S&=&\int d^3 x\, \left[\bar{f}_L\, \sigma^\mu_L ( \partial_{\mu}+i a_{\mu,L}) f_L+\bar{f}_R\, {\sigma}^\mu_R ( \partial_{\mu}+i a_{\mu,R}) f_R\right] \\ 
\label{eq:S3s2b}
\nonumber
&-&\frac{i}{4 \pi} \mathrm{sgn}(m_L) a_Ld A_L -\frac{i}{8 \pi} \mathrm{sgn}(m_L) a_Lda_L\\
&-&\frac{i}{4 \pi} \mathrm{sgn}(m_R) a_Rd A_R -\frac{i}{8 \pi} \mathrm{sgn}(m_R) a_Rda_R+\cdots.
\end{eqnarray}
Each sector is an instance of Son's duality. Because we kept track of $m_{\chi}$, the dependence on the sign of the mass of the mutual Chern-Simons term in this construction is explicit, and signals the choice related to the presence or absence of a finite Hall conductivity~\cite{Senthil:2019tl,Agarwal:2019cf}. 

Depending on the relative sign of $m_L$ and $m_R$, we can arrive at two different dualities. Physically, the different sign choices represent different realizations of time-reversal symmetry, as discussed after Eq.~\eqref{CSeff}. If they are equal, we find, using the relations in Appendix \ref{app:useful}, the dual theory
\begin{subequations}
\label{eq:cQED32}
\begin{eqnarray}
\label{eq:cQED3a2}
\nonumber
S^{(1)}&=&\int d^3 x\, \bar{f} \gamma^\mu ( \partial_{\mu}+i a_{\mu}+i\gamma_5 a_{5,\mu}) f \\
&-& \dfrac{i}{2\pi}adA - \dfrac{i}{2\pi}a_{5}d A_{5}- \dfrac{i}{4\pi}ada - \dfrac{i}{4\pi}a_{5}d a_{5}+\cdots.\nonumber\\
\end{eqnarray}
If the signs of $m_L$ and $m_R$ are opposite, the dual theory is
\begin{eqnarray}
\label{eq:cQED3b2}
\nonumber
S^{(2)}&=&\int d^3 x\, \bar{f} \gamma^\mu ( \partial_{\mu}+i a_{\mu}+i\gamma_5 a_{5,\mu}) f \\
&-& \dfrac{i}{2\pi}adA_5 - \dfrac{i}{2\pi}a_{5}d A- \dfrac{i}{2\pi}ad a_{5}+\cdots.
\end{eqnarray}
\end{subequations}
After replacing $a\to -a$ and $a_5\to -a_5$, we obtain the dualities given in Eqs.~\eqref{eq:cQED3i}. 
The dualities between Eqs.~\eqref{eq:Diracall} and~\eqref{eq:cQED3i} are the main result of this section. They encompass the generalization of the fermion-fermion duality~\cite{Son:2015gz} in the presence of axial fields.

\subsection{Effective actions in the massive case}

Let us now check the equivalence between the effective actions that result from Eqs.~\eqref{eq:Diracall} and~\eqref{eq:cQED3i} when a mass term is added. It is sufficient to focus on a single sector of Eqs.~\eqref{eq:Diracall}; we choose $\chi=L$ as the derivation for $\chi=R$ is analogous. For the $\psi$ fermions, by adding an arbitrary mass term $m_A\bar{\psi}_L\psi_L$ we obtain the effective action~\cite{Niemi:1983hw}:
\begin{eqnarray}\label{CS3}
S_{A,\mathrm{eff}}&=&\frac{i}{8 \pi} \int d^3 x\, \left(\mathrm{sgn}(m_A)+\mathrm{sgn}(m_{L})\right) A_{L}dA_{L}.
\end{eqnarray}
For the $f$ fermions, we can similarly add a mass term $m_B\bar{f}_L f_L$ to Eq.~\eqref{eq:S3s2} and integrate out the fermions to obtain
\begin{eqnarray}\label{CS4}
S_{B,\mathrm{eff}}&=&\int d^3 x\,  \frac{i}{8 \pi}(\mathrm{sgn}(m_B)-\mathrm{sgn}(m_L)) a_{L}d a_{L}\nonumber \\
&-&i\frac{\mathrm{sgn}(m_L)}{4 \pi} a_Ld A_L.
\end{eqnarray}
Following the steps outlined in Appendix \ref{app:intout}, we can integrate out the field $a_L$ to obtain
\begin{eqnarray}
    S_{B,\mathrm{eff}}&=&\frac{i}{8 \pi} \int d^3 x
    (-\mathrm{sgn}(m_B)+\mathrm{sgn}(m_L))A_{L}dA_{L},
\end{eqnarray}
which coincides with Eq.~\eqref{CS3} if we identify $m_A=-m_B$. This identification implies that the mass term has opposite signs on opposite sides of the duality, recovering a known property of fermion-fermion dualities~\cite{Metlitski,Sachdev:qQdEQuFt}.

\section{3+1 Duality Fermion-Fermion duality with an axial gauge field \label{sec:3+1}}

\subsection{Formulation of the duality\label{sec:form3+1}}
Here we extend the axial slave-rotor approach to connect two theories involving massless Dirac fermions in 3+1 dimensions, each composed of two Weyl fermions with opposite chiralities. 
Our 3+1-dimensional duality connects a free Dirac fermion coupled to external vector ($A_\mu$) and axial ($A_{\mu,5}$) fields, given by
\begin{eqnarray}
\label{eq:W1}
    S_{c_1}&=& \int d^4x \bar{\psi} \gamma^\mu(\partial_\mu-i A_\mu-i\gamma_5A_{5,\mu})\psi,
\end{eqnarray}
in Euclidean space, to an interacting theory with two emergent fields U(1) fields, a vector ($a_\mu$) and an axial field ($a_{5,\mu}$), and two Kalb-Ramond fields ($B_{\mu\nu}$ and $B_{5,\mu\nu}$) that read
\begin{eqnarray}
\label{eq:W2}
\nonumber
    S_{c_2}&=& \int d^4 x\, \bar{f} \gamma^\mu (\partial_{\mu}+i a_{\mu}+i a_{5,\mu}\gamma_5) f  \\
    \nonumber
    &-&i\epsilon^{\mu\nu\rho\sigma}[ F_{\mu\nu}(A+a)B_{\rho\sigma}+ F_{\mu\nu}(A_5+a_5)B_{5,\rho\sigma}]+\\
    \nonumber
    &+&  \left. \frac{i}{4\pi^2}\epsilon^{\mu\nu\rho\sigma}a_{5,\mu} \left(A_{\nu}\partial_\rho A_{\sigma}+A_{5,\nu}\partial_\rho A_{5,\sigma}\right)\right.\\
    &+& \frac{i}{4\pi^2}\epsilon^{\mu\nu\rho\sigma}a_{\mu} \left(A_{5,\nu}\partial_\rho A_{\sigma}+A_{\nu}\partial_\rho A_{5,\sigma}\right)+\cdots,
\end{eqnarray}
where $F_{\mu\nu}(A+a)=\partial_\mu (A_\nu+a_\nu)-\partial_\nu (A_\mu+a_\mu)$.
This duality reduces to that derived in Ref.~\cite{Palumbo2019} once the external and emergent axial fields are switched off. It includes the particularly interesting case when $A_{5,\mu}$ is chosen to be a constant, $-b_\mu$. This theory breaks Lorentz invariance~\cite{Colladay1997,Colladay1998}, and describes a Weyl semimetal with two nodes separated in momentum space by $2b_\mu$ [see Fig.~\ref{fig:3+1}(b)]~\cite{Zyuzin2012,Grushin12,Goswami12}. It reads
\begin{eqnarray}
\label{eq:W1b}
    S_{W_1}&=& \int d^4x\bar{\psi} \gamma^\mu(\partial_\mu+i A_\mu+i\gamma_5b_{\mu})\psi.
\end{eqnarray}
We find that its dual theory is given by
\begin{eqnarray}
\label{eq:W2b}
\nonumber
    S_{W_2}&=& \int d^4 x\, \bar{f} \gamma^\mu (\partial_{\mu}+i a_{\mu}+i b_{\mu}\gamma_5) f  \\
    \nonumber
    &-&i\epsilon^{\mu\nu\rho\sigma}F_{\mu\nu}(a-A)B_{\rho\sigma}\\
    &-&\dfrac{i}{4\pi^2}\epsilon^{\mu\nu\rho\sigma}b_\mu [a_\nu\partial_{\rho}(A_\sigma+a_\sigma) -2 A_\nu\partial_{\rho}A_\sigma]+\cdots.
\end{eqnarray}
These are the main results of this section, and are summarized in Fig.~\ref{fig:3+1}.

\subsection{Derivation of the duality}

The derivation of the 3+1-dimensional duality proceeds similarly to the one in the previous section.
The main difference is the role played by the chiral anomaly, which is relevant for massless fermions in even spacetime dimensions.

We start by considering a three-dimensional tight-binding model on a cubic lattice for fermions coupled to an external electromagnetic field $A_\mu$ and an axial gauge field $A_{5,\mu}$. The corresponding Hamiltonian is given by
\begin{eqnarray}\label{Dirac3}
H_{\psi}=\sum_{r,s}\left[\psi^{\dagger}_{r}\left(\frac{-m\gamma^{0}+\gamma^{0} \gamma^{s} }{2}\right)e^{-i\left( A_{r,r+\hat{s}}+\gamma_5 A_{5,r,r+\hat{s}}\right)}\psi_{r+\hat{s}} \right] \nonumber\\
+{\rm h.c.}+\sum_{r}\psi^{\dagger}_{r}\left[ 3m\gamma^{0}-i A_{r,0}-i \gamma_5 A_{5,r,0}\right]\psi_{r},\hspace{1.3cm}
\end{eqnarray}
where $r=\{x,y,z\}$ is the site index,  $A_{r,r+\hat{s}}$ and $A_{5,r,r+\hat{s}}$ are introduced through a Peierls substitution on the lattice link $(r, r+\hat{s})$ with $\hat{s}\equiv (\hat{x},\hat{y}, \hat{z})$,  $\psi_{r}$ is a four-component spinor and $\gamma^{\mu}$ are the Dirac matrices in the (Euclidean) chiral basis, defined as $\gamma^x= -\sigma_y \otimes \sigma_x$, $\gamma^y= -\sigma_y \otimes \sigma_y$, $\gamma^z= -\sigma_y \otimes \sigma_z$, $\gamma^0=\sigma_x \otimes \mathbb{I}$, with $\gamma^{5}=-\gamma^{0}\gamma^{x}\gamma^{y}\gamma^{z}=-\sigma_{z}\otimes \mathbb{I}$ the chiral matrix. 

This Hamiltonian interpolates between different topological phases depending on the value of the parameters and the configuration of the gauge fields. For example, upon choosing constant components of the chiral gauge field ($A_{5,r,0}=b_0$ and $A_{5,r,r+\hat{s}} \equiv b_{\hat{s}}$) and expanding the exponential that contains the latter to first order, this model realizes the Hamiltonian in Ref.~\cite{Behrends2019}. It features a Weyl semimetal and topological insulator phases, depending on the relative magnitude of $-b^2=b^2_{0}-\mathbf{b}^2$ and $m^2$ (see Refs.~\cite{Vazifeh2013,Grushin15,Behrends2019} for a discussion). 

The imaginary-time action corresponding to Hamiltonian \eqref{Dirac3} can be written as follows:
\begin{widetext}
\begin{eqnarray}\label{action0}
S&=&\int_{0}^{\beta}d\tau \sum_{r,s}\left[\psi_{r,L}^{\dagger}\partial_{\tau} \psi_{r,L}+\psi_{r,R}^{\dagger}\partial_{\tau} \psi_{r,R} -i n_{r,L}(A_{r,0}+
A_{5,r,0})-i n_{r,R}(A_{r,0}-A_{5,r,0})  \right.
\nonumber \\ 
\nonumber
&+&\left.\psi_{r,L}^{\dagger}\dfrac{\sigma^{s}}{2}e^{-i( A_{r,r+\hat{s}}+A_{5,r,r+\hat{s}})}\psi_{r+\hat{s},L}
-\psi_{r,R}^{\dagger}\dfrac{\sigma^{s}}{2}e^{-i( A_{r,r+\hat{s}}-A_{5,r,r+\hat{s}})}\psi_{r+\hat{s},R}\right.\\
&+&\left. 3 m \left(\psi^{\dagger}_{r,L}\psi_{r,R}+ \psi^{\dagger}_{r,R}\psi_{r,L}\right)
-\dfrac{m}{2}\left(\psi_{r,L}^{\dagger}e^{-i( A_{r,r+\hat{s}}-A_{5,r,0})}\psi_{r+\hat{s},R}
+\psi_{r,R}^{\dagger}e^{-i( A_{r,r+\hat{s}}+A_{5,r,0})}\psi_{r+\hat{s},L}
+{\rm h.c.}\right)\right].
\end{eqnarray}
The terms proportional to $m$ mix both chiralities. Upon choosing $m=0$ and expanding close to the $\Gamma$ point, the low-energy action realizes a massless Dirac fermion coupled to two gauge fields~\cite{Behrends2019}:
\begin{eqnarray}\label{action00t}
S=\int d^4 x\, \bar{\psi} \gamma^\mu (\partial_{\mu}-i A_{\mu}-i A_{5,\mu}\gamma_5) \psi 
=\int d^4 x\, \left[\bar{\psi}_L\, \sigma^\mu_L (\partial_{\mu}-i A_{\mu,L}) \psi_L\right.
+\left.\bar{\psi}_R\, \sigma^\mu_R (\partial_{\mu}-i A_{\mu,R}) \psi_R\right], 
\end{eqnarray}
with $\sigma^\mu_L=(\mathbb{I}, \sigma^i)$, and ${\sigma}^\mu_R=(\mathbb{I}, -\sigma^i)$.
As in the 2+1-dimensional case, we employ the axial slave-rotor approach to derive the dual theory of Eq.~\eqref{action00t}. As in the 2+1-dimensional case we are interested in the massless limit, and so we again neglect terms proportional to $m$ by choosing this parameter to be small. 
By substituting Eq.~(\ref{new}) in the action (\ref{action0}), we have
\begin{eqnarray}
\label{action1}
\nonumber 
S&=&\int_{0}^{\beta}d\tau \sum_{r,s}
\left[f_{r,L}^{\dagger}\partial_{\tau} f_{r,L}+f_{r,R}^{\dagger}\partial_{\tau} f_{r,R}
-i n_{r,L}(\partial_{\tau}\theta_{r,L}+A_{r,0,L})-i n_{r,R}(\partial_{\tau}\theta_{r,R}+A_{r,0,R}) \right.\\
\nonumber 
 &+&\left. i \lambda_{r,L}(f_{r,L}^{\dagger}f_{r,L}-n_{r,L}-1)+i\lambda_{r,R}(f_{r,R}^{\dagger}f_{r,R}-n_{r,R}-1)\right.\\
&+&\left.f_{r,L}^{\dagger}\left(\sigma^{s} /2\right)e^{-i( A_{r,r+\hat{s},L}+\Delta_s \theta_{r,L})}f_{r+\hat{s},L}+f_{r,R}^{\dagger}\left(-\sigma^{s} /2\right)e^{-i( A_{r,r+\hat{s},R}+\Delta_s \theta_{r,R})}f_{r+\hat{s},R}+{\rm h.c.}\right].  
\end{eqnarray}
\end{widetext}
As before, $\lambda_{r,L}$ and $\lambda_{r,R}$ are the Lagrange multiplier fields that impose the constraints Eq.~\eqref{eq:constraint}.
To decouple the rotor field and the gauge field from the fermions in the terms in the third row in Eq.~(\ref{action1}) we introduce two Hubbard-Stratonovich fields $h_L\equiv\zeta_L e^{i a_L}$ and $h_R\equiv\zeta_R e^{i a_R}$ defined on the lattice. By considering their magnitudes $\zeta_{L/R}$ constant, Eq.~(\ref{action1}) can be rewritten as follows:
\begin{eqnarray}
\label{action2}
\nonumber
S&=&\int_{0}^{\beta}d\tau \sum_{r,s,\chi=L,R}\left[f_{r,\chi}^{\dagger}(\partial_{\tau}+i a_{r,0,\chi}) f_{r,\chi}\right.\\
&-& \left. i n_{r,\chi}(\partial_{\tau}\theta_{r,\chi}+A_{r,0,\chi}+a_{r,0,\chi})\nonumber\right.\\
\nonumber
&+&\left.\hat{\chi}\left(\zeta_\chi f_{r,\chi}^{\dagger}(\sigma^{s}/2)e^{i a_{r,r+\hat{s},\chi}}f_{r+\hat{s},\chi}+{\rm h.c.}
\right)\right. \\
&-&\left.\zeta_\chi \cos \left(\Delta_{\hat{s}}\theta_{r,\chi}+A_{r,r+\hat{s},\chi}+a_{r,r+\hat{s},\chi}\right)\right],
\end{eqnarray}
where we have reinstated the notation that the scalar $\hat{\chi}$ takes the value $\hat{\chi}=+1$ and $\hat{\chi}=-1$ for chiralities $\chi=L$  and $\chi=R$, respectively. Similar to the 2+1-dimensional case, we have defined $a_{r,0,\chi}\equiv \lambda_{r,\chi}$. After employing the Villain approximation for the last two term terms in the above equation, the action can be decomposed in two terms,
\begin{eqnarray}
\label{eq:total}
S=S_{f}+S_{\theta},
\end{eqnarray}
where
\begin{eqnarray}\label{action3}
\nonumber
S_f&=&\int_{0}^{\beta}d\tau \sum_{r,s,\chi=L,R}\left[f_{r,\chi}^{\dagger}(\partial_{\tau}+i a_{r,0,\chi}) f_{r,\chi}\right.\\
&+& \left.\hat{\chi}\left(\zeta_\chi f_{r,\chi}^{\dagger}(\sigma^{s}/2)e^{i a_{r,r+\hat{s},\chi}}f_{r+\hat{s},\chi}+{\rm h.c.}\right)
\right],
\end{eqnarray}
and
\begin{eqnarray}\label{S2}
S_{\theta}&=&\int_{0}^{\beta}d\tau \sum_{r,s,\chi=L,R}\left[i J^{r,0}_\chi(\partial_{\tau}\theta_{r,\chi}+A_{r,0,\chi}+a_{r,0,\chi})+  \right. \nonumber \\ 
\nonumber
&+&\left. i J^{r,r+\hat{s}}_\chi \left(\Delta_{\hat{s}}\theta_{r,\chi}+A_{r,r+\hat{s},\chi}+a_{r,r+\hat{s},\chi}\right)\right.\\
&+&\left.\dfrac{1}{2\zeta_\chi}\left(J^{r,r+\hat{s}}_\chi\right)^{2} \right],
\end{eqnarray}
where we have identified $n_{r,\chi}$ as the temporal component, $J^{r,0}_\chi$, of the bosonic current $J^{r,r+\hat{s}}_\chi$. In the long-wavelength limit, Eq.~\eqref{action3} becomes
\begin{eqnarray}
S_{f}=\int  d^{4}x\, \bar{f}_L\, \sigma^{\mu}_L(\partial_{\mu}+ia_{\mu,L})f_L+\bar{f}_R\, \sigma^{\mu}_R(\partial_{\mu}+ia_{\mu,R})f_R, \nonumber \\
\end{eqnarray}
where $\sigma^\mu_L=(\mathbb{I}, \sigma^i)$ and $\sigma_R^\mu=(\mathbb{I}, -\sigma^i)$. 
In this limit, $\Delta_{\hat{s}}$ reduces to the standard spatial derivative, and by integrating out $\theta_L$ and $\theta_R$ in Eq. (\ref{S2}), we obtain
\begin{eqnarray}
\label{eq:currentcons}
\partial_{\mu}J^{\mu}_\chi=0. 
\end{eqnarray}
A solution for these two equations is given by
\begin{eqnarray}
J^{\mu}_\chi=\epsilon^{\mu\nu\lambda\delta}\partial_{\nu}B_{\lambda\delta,\chi}, 
\end{eqnarray}
where $B_{\lambda\delta, L}$ and $B_{\lambda\delta, R}$ are antisymmetric tensor (Kalb-Ramond) gauge fields. 

At this point, it is important to recall that in 3+1 dimensions the path integral measure is not invariant under the transformations Eq.~\eqref{new}, a fact known as the chiral anomaly~\cite{B96}. Therefore, there is an additional contribution to the effective action that takes into account the non-conservation of chiral charge. It is of the form~\cite{B96,Fujikawa1984}
\begin{eqnarray}
\label{eq:anomalies}
    S_{\mathrm{an}} &=& i\int d^4x \hspace{1mm} \theta_{\chi}\mathcal{A}_{\chi}(x),
\end{eqnarray}
where $\mathcal{A}_{\chi}(x) = \frac{\hat{\chi}}{32\pi^2}  \epsilon^{\mu\nu\rho\sigma}F_{\mu\nu}(A_{\chi}) F_{\rho\sigma}(A_{\chi})$. This factor carries through our derivation modifying the current conservation equation Eq.~\eqref{eq:currentcons} to
\begin{eqnarray}
    \partial_\mu J_\chi^{\mu} -\frac{\hat{\chi}}{32\pi^2}\epsilon^{\mu\nu\rho\sigma}F_{\mu\nu}(A_{\chi}) F_{\rho\sigma}(A_{\chi})=0.
\end{eqnarray}
Consequently, the most general form of the current is
\begin{eqnarray}
J^{\mu}_\chi=\epsilon^{\mu\nu\lambda\delta}\partial_{\nu}B_{\lambda\delta,\chi}
+\frac{\hat{\chi}}{16\pi^2}\epsilon^{\mu\nu\rho\sigma}A_{\nu,\chi} F_{\rho\sigma}(A_{\chi}),
\end{eqnarray}

Inserting this current back into $S_\theta$, we can write Eq.~\eqref{eq:total} as
\begin{eqnarray}
\nonumber
\label{eq:WeylLR}
    S&=& S_f+\int  d^{4}x\,\sum_{\chi=L,R}\epsilon^{\mu\nu\rho\sigma} i\partial_\nu B_{\rho\sigma,\chi}(A_{\mu,\chi}+a_{\mu,\chi})\\
    &+&i\frac{\hat{\chi}}{8\pi^2}\epsilon^{\mu\nu\rho\sigma}a_{\mu,\chi}A_{\nu,\chi} \partial_{\rho}A_{\sigma,\chi},
\end{eqnarray}
where we have omitted the $J^2$ term and used the fact that $\epsilon AAdA$ identically vanishes.
By combining now the two chiralities into a compact notation, we reach the final form of the duality,
\begin{eqnarray}\label{eq:dualWeyl}
\nonumber
    S&=& \int d^4 x\, \bar{f} \gamma^\mu (\partial_{\mu}+i a_{\mu}+i a_{5,\mu}\gamma_5) f  \\
    \nonumber
    &-&i\epsilon^{\mu\nu\rho\sigma} [F_{\mu\nu}(A+a)B_{\rho\sigma}+ F_{\mu\nu}(A_5+a_5)B_{5,\rho\sigma}]\\
    \nonumber
     &+&  \left. \frac{i}{4\pi^2}\epsilon^{\mu\nu\rho\sigma}a_{5,\mu} \left(A_{\nu}\partial_\rho A_{\sigma}+A_{5,\nu}\partial_\rho A_{5,\sigma}\right)\right.\\
     \nonumber
    &+& \frac{i}{4\pi^2}\epsilon^{\mu\nu\rho\sigma}a_{\mu}\left(A_{5,\nu}\partial_\rho A_{\sigma}+A_{\nu}\partial_\rho A_{5,\sigma}\right)+\cdots,\\
\end{eqnarray}
where $B_{\rho\sigma}=(B_{\rho\sigma,L}+B_{\rho\sigma,R})/2$ and $B_{5,\rho\sigma}=(B_{\rho\sigma,L}-B_{\rho\sigma,R})/2$.
The two last lines ensure that the anomaly is the same on both sides of the duality, as we show in the next section. 

Finally, we remark that the kinematic terms for the Kalb-Ramond fields, hidden in $+\cdots$, prevent the theory from gapping out due to the existence of dynamical string-like excitations, similar to Ref.~\cite{Palumbo2019}. This is compatible with the absence of condensation of the slave rotor field that prevent the formation of a Mott insulating phase. This is an assumption that is inherent to this approach, as anticipated in Section~\ref{sec:rotor}. 

\subsection{Consistency of the chiral anomaly\label{eq:anomalycheck}}
As discussed above, the gapless $\psi$ fermions are anomalous, which implies that the combined vector and axial gauge transformations of Eq.~\eqref{eq:W1} result in the effective action~\cite{B96},
\begin{eqnarray}\label{ST1}
\nonumber
S^{(1)}_{\theta}&=& \int d^4 x\, \bar{\psi} \gamma^\mu (\partial_{\mu}-i A_{\mu}
    -i A_{5,\mu}\gamma_5) \psi  \\
    &+&i\theta(\partial_\mu J^{\mu}+\frac{\epsilon^{\mu\nu\lambda\delta}}{8\pi^2} F_{\mu\nu}(A)F_{\lambda\delta}(A_5))\nonumber\\
\nonumber
&+& i\theta_5(\partial_\mu J^{\mu}_5+\frac{\epsilon^{\mu\nu\lambda\delta}}{16\pi^2}\left[ F_{\mu\nu}(A)F_{\lambda\delta}(A)\right.\\
&+&\left.F_{\mu\nu}(A_5)F_{\lambda\delta}(A_5)\right]),
\end{eqnarray}
where $\theta$ and $\theta_5$ are related to $\theta_\chi$ in Eq.~\eqref{eq:anomalies} by the relation $\theta_{\chi}= \theta +\hat{\chi}\theta_5$. This formulation of the anomaly, known as the covariant anomaly, might look worrisome, since the vector current is not explicitly conserved ($\partial_\mu J^{\mu}\neq 0$).  This problem is fixed by additional current terms known as Bardeen polynomials, which impose gauge invariance and define the consistent anomaly that explicitly conserves the vector current~\cite{B96,Behrends2019}. For our purposes, it is enough to set aside this issue and work with the covariant anomaly, keeping in mind that it has a standard solution.

By construction, the $f$ fermion side of the duality, Eq.~\eqref{eq:dualWeyl}, also contains the same chiral anomaly. 
By varying Eq.~\eqref{eq:dualWeyl} with respect to the Kalb-Ramond fields $B_{\mu\nu}$ and $B_{5,\mu\nu}$, we arrive at the constraints
\begin{eqnarray}\label{ST3}
F_{\lambda\delta}(A_\chi)+F_{\lambda\delta}(a_\chi)=0, 
\end{eqnarray}
implying that $a_\chi= - (A_{\mu,\chi} +\partial_\mu \xi_\chi)$. By inserting these expressions back into Eq.~\eqref{eq:dualWeyl}, we obtain
\begin{eqnarray}
\nonumber
 S^{(2)}_{\theta}&=& \int d^4 x\, \bar{f} \gamma^\mu (\partial_{\mu}-i A_{\mu}
    -iA_{5,\mu}\gamma_5) f  \\
    \nonumber
    &+&i\xi(\partial_\mu J^{\mu}+\frac{\epsilon^{\mu\nu\lambda\delta}}{8\pi^2} F_{\mu\nu}(A)F_{\lambda\delta}(A_5))\nonumber\\
\nonumber
&+&i\xi_5(\partial_\mu J^{\mu}_5+\frac{\epsilon^{\mu\nu\lambda\delta}}{16\pi^2}\left[ F_{\mu\nu}(A)F_{\lambda\delta}(A)\right.\\
&+&\left.F_{\mu\nu}(A_5)F_{\lambda\delta}(A_5)\right]).
\end{eqnarray}
This shows that both theories have the same anomaly as $S^{(1)}_{\theta}$ if we identify $\theta=\xi$ and $\theta_5=\xi_5$. 
Although obtaining the same anomaly is a consistency check, it is to some extent not surprising. Our generalized slave-rotor approach, and in particular Eq.~\eqref{eq:dualWeyl}, was built to incorporate the same chiral anomaly on both sides of the duality. In the next section, we study a specific case of our duality, which concerns the theory of a Weyl semimetals, and gives us a nontrivial consistency check of our results.

\subsection{Weyl duality and connection to the 2+1-dimensional fermionic duality \label{sec:Weylduality}}
In this section we derive a duality between two Weyl semimetal theories. In particular, we wish to derive the dual to
\begin{eqnarray}
	\label{eq:LBQEDap}
	S_b&=&\int d^4x\; \bar{\psi}\gamma^{\mu}(\partial_\mu + iA_\mu +ib_\mu\gamma_5)\psi.
	\end{eqnarray}
As discussed extensively in the literature (see, for example, Refs.~\cite{Grushin12,Grushin15,Zyuzin2012,Goswami12,Ramamurthy15}), this theory describes two Weyl fermions separated in energy-momentum space by $2 b_\mu$. The vector $b_\mu$ is a constant vector in space-time, and thus breaks Lorentz symmetry~\cite{Colladay1997,Colladay1998}.

A chiral transformation, where  $\bar{\psi}\to \bar{\psi} e^{i\theta_5(x)\gamma_5}$  and $\psi\to e^{i\theta_5(x)\gamma_5}\psi$, can remove $b_\mu$ from the fermionic action provided we choose $\partial_\mu\theta_5 =-b_\mu$. This transformation removes $b_\mu$ from Eq.~\eqref{eq:LBQEDap}, but adds the following Carroll-Field-Jackiw~\cite{Carroll90} term to the effective action~\cite{Chung99,Grushin12,Zyuzin2012,Goswami12}
\begin{eqnarray}
\label{eq:Weyleffb}
    S_{b} = -\dfrac{i}{4\pi^2} \int d^4 x\, \epsilon^{\mu\nu\rho\sigma} b_\mu A_\nu \partial_\rho A_\sigma.
\end{eqnarray}
Rotating back to real time results in the electromagnetic current 
\begin{eqnarray}
\label{eq:currents}
    J^{\mu} = \dfrac{\delta S}{\delta A_\mu} = \dfrac{1}{2\pi^2} \epsilon^{\mu\nu\rho\sigma} b_\nu \partial_\rho A_\sigma,
\end{eqnarray}
which describes, for example, the quantum Hall effect proportional to the Weyl node separation, a known characteristic of the Weyl semimetal phase~\cite{Burkov:2011de}.

To derive the dual of Eq.~\eqref{eq:LBQEDap} from Eq.~\eqref{eq:dualWeyl}, we notice that the field $B_{5,\mu\nu}$ acts as a Lagrange multiplier by neglecting the higher-order kinetic terms $\propto (\partial B_5)^2$. Then, by integrating $B_{5,\mu\nu}$ out, we obtain the condition $a_{\mu,5}=-(A_{5,\mu}+\partial_\mu \xi_5)$. Inserting this condition in Eq.~\eqref{eq:dualWeyl}, and noting that $A_\mu$ in Eq.~\eqref{eq:LBQEDap} enters with an opposite sign with respect to our original theory Eq.~\eqref{eq:W1} we arrive at
\begin{eqnarray}
\label{eq:thisweyl}
\nonumber
    S&=& \int d^4 x\, \bar{f} \gamma^\mu (\partial_{\mu}+i a_{\mu}-i (A_{5,\mu}+\partial_\mu\xi_5)\gamma_5) f  \\
    \nonumber
    &-&i\epsilon^{\mu\nu\rho\sigma}F(-A_\mu+a_\mu)B_{\rho\sigma}\\
    \nonumber
   &-&  \left. \frac{i}{4\pi^2}\epsilon^{\mu\nu\rho\sigma}(A_{5,\mu}+\partial_\mu \xi_5 )\left(A_{\nu}\partial_\rho A_{\sigma}+A_{5,\nu}\partial_\rho A_{5,\sigma}\right)\right.\\
    &-& \frac{i}{4\pi^2}\epsilon^{\mu\nu\rho\sigma}a_{\mu}\left(A_{5,\nu}\partial_\rho A_{\sigma}+A_{\nu}\partial_\rho A_{5,\sigma}\right)+\cdots.
\end{eqnarray}
From the anomaly matching in the last section, we identify $\partial_\mu \xi_5 = \partial_\mu \theta_5 = -b_\mu$, and from Eq.~\eqref{eq:LBQEDap} we can read off that $A_{5,\mu} = -b_\mu$, leading to
\begin{eqnarray}
\label{eq:thisweyl2}
\nonumber
    S&=& \int d^4 x\, \bar{f} \gamma^\mu (\partial_{\mu}+i a_{\mu}+2ib_\mu) f -i\epsilon^{\mu\nu\rho\sigma}F_{\mu\nu}(a-A)B_{\rho\sigma}\\
    &+&  \left. \frac{i}{4\pi^2}\epsilon^{\mu\nu\rho\sigma}b_\mu(2A_{\nu}-a_\nu)\partial_\rho A_{\sigma}\right. .
\end{eqnarray}
%
To bring it to a more recognizable form, we now choose to perform a chiral transformation to remove one $b_\mu$ from the first term, adding a term like Eq.~\eqref{eq:Weyleffb} to the effective action, but with $A_\mu$ replaced by $a_\mu$. This results in
\begin{eqnarray}
\label{eq:weylfinal}
\nonumber
    S&=& \int d^4 x\, \bar{f} \gamma^\mu (\partial_{\mu}+i a_{\mu}+i b_{\mu}\gamma_5) f  -i\epsilon^{\mu\nu\rho\sigma}F_{\mu\nu}(a-A)B_{\rho\sigma}\\
    &-&\dfrac{i}{4\pi^2}\epsilon^{\mu\nu\rho\sigma}b_\mu [a_\nu\partial_{\rho}(A_\sigma+a_\sigma)- 2A_\nu\partial_{\rho}A_\sigma]+\cdots.
\end{eqnarray}
%
This is our final form for the dual action, and we now ask if it recovers Eq~\eqref{eq:LBQEDap} and, consequently, Eq.~\eqref{eq:currents}.
As before, we may integrate out $B_{\rho\sigma}$, which in this case imposes that $a_\mu = A_\mu+\partial_\mu\zeta$. Inserting it into Eq.~\eqref{eq:weylfinal} the terms with  $\partial_\mu\zeta$ drop out, and the last two rows cancel, resulting in the effective action Eq.~\eqref{eq:LBQEDap}, but with the replacement $b \to -b$. 

We find that the sign change that maps $b$ to $-b$ is implied by Son's 2+1-dimensional duality. To see this, we recall that the Weyl theory in Eq.~\eqref{eq:LBQEDap} can be viewed as a collection of 2+1-dimensional massive Dirac theories with masses parametrized by the momentum along the Weyl node separation~\cite{Burkov:2011de,Wan2011} (see also Appendix ~\ref{app:Soncheck}). The points where the mass vanishes set the location of the two Weyl nodes.
Each 2+1-dimensional theory is independently subject to Son's duality, which requires the masses to change sign~\cite{Metlitski,Sachdev:qQdEQuFt}. As we describe in detail in Appendix ~\ref{app:Soncheck}, inverting the sign of the masses of the 2+1-dimensional Dirac theories results in a Hall conductivity where $b\to -b$, consistent with what we observe in our Weyl duality.
If we had obtained the same response at both sides of the Weyl duality, it would have contradicted how the mass enters in Son's fermion-fermion duality. Therefore, the mapping $b\to -b$ acts as a consistency check of our Weyl duality.

\section{Physical implications of axial gauge field dualities \label{sec:physics}}

Axial gauge fields exist in different physical systems ranging from condensed matter to high-energy physics. In this section, we discuss the implications of our 2+1- and 3+1-dimensional dualities for several condensed matter systems: 2D surfaces of 3D non-symmorphic topological insulators, 3D Weyl semimetals, and the 3D Hall effect.

\subsection{Surfaces of 3D non-symmorphic topological insulators}

In 2+1 dimensions, axial gauge fields can emerge in 2D materials like graphene~\cite{Seradjeh2008,Cortijo10,Marzuoli_2012,amorim20161}, but also at the surface of 3D non-symmorphic Dirac insulators~\cite{Wieder246}, where our duality finds special significance. Non-symmorphic Dirac insulators are three-dimensional insulators with two non-symmorphic glide symmetries that topologically protect a doubly degenerate Dirac cone at the surface. The surface theory is described by a $4\times 4$ Dirac Hamiltonian, i.e. two copies of the surface state of a time-reversal symmetric topological insulator~\cite{Kane}. This effective theory naturally allows us to introduce an axial gauge field that couples with opposite signs to each copy. Similar to graphene, this axial gauge field arises from the presence of strain at the boundary of the non-symmorphic topological insulator. 

Son's original 2+1-dimensional duality suggested the existence of a dual theory of the surface of a 3D time-reversal invariant topological insulator~\cite{Metlitski,CWang}. In a similar way, our 2+1-dimensional duality suggests that the boundary of strained non-symmorphic topological insulators has a dual metallic boundary phase characterized by an emergent neutral fermion $f$ coupled to two emergent gauge fields, described by Eqs.~\eqref{eq:cQED3i}. The existence of the axial field is crucial for these theories, differentiating them from a simple doubling of Son's dual theory. They therefore suggest the existence of a dual strain-induced critical phase for the surface of 3D non-symmorphic topological insulators. 

It may be possible to explicitly show the duality between surface theories in strained non-symmorphic topological insulators by extending the bulk electromagnetic duality used in Ref.~\cite{Metlitski}. Their construction viewed Son's duality as a duality between two surface theories at the surface of two dual bulk topological insulators. By incorporating bulk crystalline symmetries to this construction one could account for axial fields at the boundary, and derive a duality between surface theories with axial gauge fields. This is a possibility we leave for future work.

\subsection{Weyl semimetals and the quantized circular photogalvanic effect}

One interesting consequence of the duality between Eqs.~\eqref{eq:W1b} and \eqref{eq:W2b} concerns their non-linear responses. In Fourier space, Eq.~\eqref{eq:W1b} describes a Weyl semimetal with nodes separated both in energy and momentum space. Upon shining circularly polarized light, such a Weyl semimetal responds with an \textit{exactly} quantized circular photogalvanic effect, which is the part of the induced photocurrent that changes sign with the sense of circular polarization~\cite{deJuan17}. The photocurrent shows a frequency plateau, quantized to the Weyl monopole charge $C$ in units of $\pi e^3/h^2$. If the duality between Eq.~\eqref{eq:W1b} and Eq.~\eqref{eq:W2b} holds, then Eq.~\eqref{eq:W2b} also displays a quantized circular photogalvanic effect. 

This correspondence is important because the quantized circular photogalvanic effect is in general corrected by electron-electron interactions~\cite{Avdoshkin20}, unlike the quantized Hall conductivity of a two-dimensional insulator. The duality between Eq.~\eqref{eq:W1b} and Eq.~\eqref{eq:W2b} implies that the interactions between the neutral $f$ fermions with the gauge and Kalb-Ramond field conspire to deliver a quantized circular photogalvanic effect as a response to the external field $A_\mu$. 

Although it is tempting to regard Eq.~\eqref{eq:W2b} as the first example of an interacting theory with a quantized non-linear response, and among the few that display this effect~\cite{deJuan17,ChangEA17,Flicker2018}, it is important to be cautious. The correspondence between responses follows straightforwardly when we are allowed to integrate out the Kalb-Ramond field $B_{\rho\sigma}$. This leads to the condition $a_\mu \to A_\mu$ and the two theories and their responses map onto each other, as discussed in Sec.~\ref{sec:Weylduality}. The implications of the duality become more profound when higher-order derivative terms in $B_{\rho\sigma}$ cannot be neglected. In this case it is not obvious that Eq.~\eqref{eq:W2b} shows a quantized non-linear response, and hence the equivalence implied by the duality is more significant.

Additionally, these observations do not imply full protection from interaction corrections. If screened Coulomb or Hubbard electron-electron interactions are present (hidden in $+\cdots $), these can still correct the circular photogalvanic effect in perturbation theory~\cite{Avdoshkin20}. To be precise, our duality between Eq.~\eqref{eq:W1b} and~\eqref{eq:W2b} implies that the types of interactions that couple $A_\mu$ to $f$ fermions, the Kalb-Ramond $B_{\mu\nu}$ and statistical gauge field $a_\mu$ in  Eq.~\eqref{eq:W2b}, do not correct the quantized circular photogalvanic effect.

\subsection{3D quantum Hall effect}

The action Eq.~\eqref{eq:W1b} is also connected to a 3D quantum Hall effect by choosing the spatial part of the axial gauge field $b_{\mu}$ to be constant and equal to a half integer multiple of a reciprocal lattice vector $\nu_i =\frac{n}{2}G_i$~\cite{Grushin12,Ramamurthy15}. In this case the effective action Eq.~\eqref{eq:W1b} results in a 3D Hall conductivity $\sigma_{xy} = \frac{ne^2}{h a_G}$, where $a_G=2\pi/|\mathbf{G}|$ is the lattice constant along $G_i$~\cite{Ramamurthy15}. This Hall conductivity is that of a layered quantum Hall system, i.e., a stack of 2D Hall insulators, each with conductivity $n e^2/h$, stacked along the reciprocal real-space direction corresponding to $\mathbf{G}$. Our duality then suggests that this theory has a dual 3D Hall theory Eq.~\eqref{eq:W2b} with $b_\mu$ replaced by $(0,\nu_i)$.

For it to be a duality between 3D Hall insulators, we have to consider the possible mechanisms that can gap out the theories at both sides of the duality. Recently, Ref.~\cite{thakurathi2020theory} proposed a possible route via a hydrodynamic BF field theory of a 3D fractional quantum Hall effect in Weyl semimetals. In this work, vortex condensation gaps out the Weyl nodes in a magnetic Weyl semimetal without breaking translational symmetry. The bosonic sector of the effective field theory describes
quasiparticles excitation that couple to an emergent and dynamical vector field $c_\mu$ and loop excitations that couple to a Kalb-Ramond field $b_{\mu\nu}$. Additionally, the statistical gauge field $a_\mu$ couples the bosonic and fermionic sectors.

Our Eq.~\eqref{eq:W2b} suggests a close connection with the theories discussed in Ref.~\cite{thakurathi2020theory}. For example, in the bosonic sector in Eq.~\eqref{eq:W2b}, we could introduce the following minimal couplings: $J^{\mu\nu} B_{\mu\nu} + J^{\mu\nu}_5 B_{5,\mu\nu}$, where $J^{\mu\nu}$ and $J^{\mu\nu}_5$ represent distinct loop currents. Together with the kinetic terms of the Kalb-Ramond fields, they describe dynamical loop currents and an eventual vortex condensation. We thus expect that combining the method in Ref.~\cite{thakurathi2020theory} with axial field dualities can lead to gapped 3D quantum Hall phases and loop excitations induced by dynamical strain that generalize those of Ref.~\cite{thakurathi2020theory}. 

\section{Discussion and conclusions \label{sec:conc}}

In this work, we have explored the role of axial gauge fields in the formulation of fermion-fermion dualities. By considering axial fields we have extended known 2+1-dimensional dualities and proposed new 3+1-dimensional dualities. They are formulated in Sections~\ref{sec:form2+1} and~\ref{sec:form3+1}, and summarized in Figs.~\ref{fig:2+1} and \ref{fig:3+1}. Our 2+1-dimensional dualities suggest the existence of dual surface theories of 3D non-symmorphic topological insulator surfaces. In 3+1-dimensions our dualities suggest that the quantization of photo-currents of Weyl semimetals is more robust than expected. They may also be used as a building block to describe gapped 3D Hall phases. 

To derive these dualities, we have extended the slave-rotor approach to include axial gauge fields. In 3+1 dimensions, this extension allows one to monitor the role of the chiral anomaly. It also has the benefit that the theories derived from it are not necessarily anisotropic. However, anisotropic methods, such as the wire~\cite{Mross,Meng}, or layered constructions~\cite{Levin,Sagi:2018en} could lead to alternative derivations of our dualities.  Additionally, an alternative and promising route to derive our 2+1 duality is to extend the bulk electromagnetic duality that applies to the 3D time-reversal-invariant topological insulators to 3D non-symmorphic topological insulators. Similarly, it may be useful to view our 3+1-dimensional duality as the boundary of a  4+1-dimensional insulator. 

However, the slave-rotor approach has known drawbacks, specifically the approximations that have been already discussed on a previous derivations of Son's duality~\cite{Burkov:2011de}. For example, the mean-field solution that we discuss is not unique since other Hubbard-Stratonovich decouplings are possible. The slave-rotor construction also relies on the absence of condensation of the rotor field or, equivalently, a Mott insulating phase. Due to the gapped nature of the Chern-Simons term, this is not an issue in 2+1 dimensions~\cite{Burkov:2011de}. In 3+1 dimensions, vortex condensation is avoided due to the existence of kinetic terms of the Kalb-Ramond fields~\cite{Palumbo2019}. Despite these limitations, the equivalence of the effective actions at both sides of the duality, and their consistency with the 2+1-dimensional fermion-fermion duality support their plausibility.

Our work shows that the known web of dualities~\cite{Seiberg:ur} could be extended to include theories with axial fields and theories with broken Lorentz invariance~\cite{Colladay1997,Colladay1998}. These types of theories seem to lie outside the focus of current duality research, despite their relevance to extensions of the standard model~\cite{Colladay1997,Colladay1998}, and topological condensed-matter systems such as Weyl semimetals~\cite{Grushin12}, nodal-line semimetals~\cite{BurkovHook2011}, and strained Dirac and Weyl systems~\cite{amorim20161,Ilan:2019io}. It is also tempting to speculate that the 3+1 duality presented in this work can be connected to a recently proposed boson-fermion duality~\cite{Furusawa:2019hx}. Lastly, the slave-rotor approach can incorporate non-Abelian gauge fields following Refs.~\cite{Hermele:2007hb,Xu:2010ga}, which may serve to derive known dualitites~\cite{Xu,Bi,Jian,Hsin,Aharony,Chen:2018cw,Argurio}, as well as novel axial non-Abelian dualities. 

Additionally, it was recently discovered that chiral semimetals can have protected band crossing with degeneracy larger than two~\cite{Manes:2012fi,bradlynScience2016,wiederPRL2016,tangPRL2017,changNatMat2018}. The excitations around these nodes, known as multifold fermions, can be described by Lorentz-breaking generalizations of Weyl fermions with monopole charge larger than one. To our knowledge, no dualities for multifold fermions exist. The slave-rotor construction can be a viable method to uncover them, both in 2+1 and 3+1 dimensions.

Finally, it is tempting to generalize our approach to higher-dimensional synthetic systems, such as 4+1-dimensional topological semimetals, where the chiral anomaly is replaced by the parity anomaly \cite{Zhu}. In this context, new three-form gauge fields $C_{\mu\nu\lambda}$ are allowed, associated to conserved bosonic currents. 

To conclude, our work emphasizes how dualities that involve axial field and Lorentz-breaking field theories can uncover the challenging phenomenology of interacting phases of gapless topological matter.  We expect that our dualities can be applied broadly beyond the condensed matter examples we used, in high-energy problems with axial gauge fields, such as the quark-gluon plasma.

\section{Acknowledgements}

A. G. G. is grateful to A. Burkov, S. Florens, F. de Juan, and S. Sayyad for discussions, and K. Driscoll for critical reading of the manuscript. G. P. acknowledges the support of the ERC through the
Starting Grant project TopoCold. A. G. G. is supported by the ANR under the grant ANR-18-CE30-0001-01 (TOPODRIVE) and the European Union Horizon 2020 research and innovation program under grant agreement No. 829044 (SCHINES). This research was supported in part by the National Science Foundation under Grant No. NSF PHY-1748958.

\newpage
\appendix
\renewcommand{\thefigure}{S\arabic{figure}}
\setcounter{figure}{0} 

\section{Some useful relations and definitions \label{app:useful}}
We list here some useful identities used in the main text.
Using that
\begin{subequations}
\begin{eqnarray}
\label{eq:fieldsapp}
a_{\mu,L}=a_{\mu}+a_{5,\mu},\\
a_{\mu,R}=a_{\mu}-a_{5,\mu},
\end{eqnarray}
\end{subequations}
the different 2+1-dimensional Chern-Simons terms can be written as follows: 
\begin{subequations}
\begin{eqnarray}
\label{eq:A1}
    a_Lda_L - a_Rda_R &=& 4ada_5, \\
\label{eq:A2}
    a_Lda_L + a_Rda_R &=& 2ada + 2a_5da_5,\\
\label{eq:A3}
    a_LdA_L - a_RdA_R &=& 2adA_5 + 2a_5dA,\\
\label{eq:A4}    
    a_LdA_L + a_RdA_R &=& 2adA + 2a_5dA_5,
\end{eqnarray}
\end{subequations}
where we have assumed it is possible to integrate by parts allowing us to identify $adA$ with $Ada$.
This latter property does not hold in 3+1 dimensions since Carroll-Field-Jackiw terms~\cite{Carroll90} like $a_5adA$ are composed of three gauge fields instead of two. Nonetheless, the following relations are useful: 
\begin{subequations}
\begin{eqnarray}
\label{eq:A1s}
    a (A_Lda_L - A_Rda_R) &=& 2a (Ada_5 + A_5da), \\
\label{eq:A2s}
    a (A_Lda_L + A_Rda_R) &=& 2a (Ada + A_5da_5).
\end{eqnarray}
\end{subequations}

\section{Effective action and mass signs in 2+1-dimensional dualities\label{app:intout}}

In this Appendix, we explicitly integrate out $a$ in Eq.~\eqref{eq:S3s} keeping track of the mass signs, which are important for our discussion, but disregarded in Ref.~\cite{Burkov2019}. We demonstrate the procedure for the left helicity, since the right helicity proceeds analogously. Defining $\slashed{D}_{a}= \sigma^{\mu}(\partial_\mu+ia_\mu)$, we write Eq. \eqref{eq:S3s} as 
\begin{widetext}
\begin{eqnarray}
    \mathcal{L}&=& \bar{f}\slashed{D}_{a}f + i\dfrac{\mathrm{sgn}(m_L)}{8\pi} ada - \dfrac{i}{4\pi}bd(A+a)\\
     &=& \bar{f}\slashed{D}_{a}f+ i\frac{\mathrm{sgn}(m_L)}{8\pi} (a-\mathrm{sgn}(m_L) b)d(a-\mathrm{sgn}(m_L) b) 
     -\dfrac{i}{4\pi}bdA -i\dfrac{\mathrm{sgn}(m_L)}{8\pi}bdb\\
     &\stackrel{\mathrm{int. out. a}}{=}& \bar{f}\slashed{D}_{b\mathrm{sgn}(m_L)}f -\dfrac{i}{4\pi}bdA-i\dfrac{\mathrm{sgn}(m_L)}{8\pi} bdb\\
     \label{eq:dualityfix}
     &\stackrel{\mathrm{sgn}(m_L) b \to a}{=}&  \bar{f}\slashed{D}_{a}f -i\dfrac{\mathrm{sgn}(m_L)}{4\pi}adA -i\dfrac{\mathrm{sgn}(m_L)}{8\pi} ada.
\end{eqnarray}
\end{widetext}
In the third line, we are allowed to integrate out $a$ because a Chern-Simons term acts like a mass term for the gauge field~\cite{Deser1982}.
When we add a mass term $m$, then we can integrate out the $f$ fermions, obtaining
\begin{eqnarray}
\label{eq:dualityfix2}
    \mathcal{L}_{\mathrm{eff}}&=& i\dfrac{\mathrm{sgn}(m)}{8\pi}ada -i\dfrac{\mathrm{sgn}(m_L)}{4\pi}adA -i\dfrac{\mathrm{sgn}(m_L)}{8\pi} ada. \nonumber \\
\end{eqnarray}
Depending on the relative sign of $m_L$ and $m$, then we can have a zero or non-zero Chern-Simons term for $a$~\cite{Senthil:2019tl}
Integrating out $a$ in Eq.~\eqref{eq:dualityfix2} implies
\begin{eqnarray}
a &=& \dfrac{\mathrm{sgn}(m_L)}{\mathrm{sgn}(m)-\mathrm{sgn}(m_L)}A.
\end{eqnarray}
Reinserting this condition into Eq.~\eqref{eq:dualityfix2} and redefining $A/(\mathrm{sgn}(m)-\mathrm{sgn}(m_L)) \to A$. we obtain
\begin{eqnarray}
\label{eq:last}
    \mathcal{L}_{\mathrm{eff}}=
    -\dfrac{(\mathrm{sgn}(m)-\mathrm{sgn}(m_L))}{8\pi}AdA.
\end{eqnarray}
This is the same Chern-Simons term we would obtain from the original theory if we identify $m$ with $-m$, at opposite sides of the duality, as expected from previous arguments~\cite{Metlitski,Sachdev:qQdEQuFt}.
\begin{figure}
    \centering
    \includegraphics[width=\columnwidth]{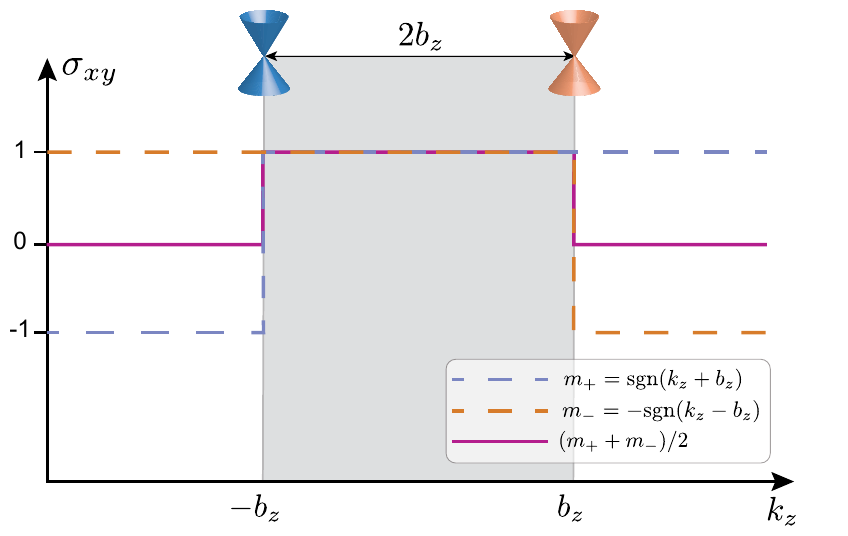}
    \caption{Hall conductivity of a theory with two Weyl cones separated in the $k_z$ direction. For each $k_z$, the theory is equivalent to two 2+1-dimensional gapped Dirac theories with $k_z$-dependent masses (dashed lines). The total 3D Hall conductivity is the integral of the solid curve, and is proportional to the Weyl node separation, see Eq.~\eqref{eq:dualWeylseffcheck1}.}
    \label{fig:hall}
\end{figure}

\section{Consistency with the 2+1-dimensional fermionic duality\label{app:Soncheck}}
We start by reminding the reader that the Weyl semimetal theory Eq.~\eqref{eq:LBQEDap}, that we repeat here for convenience 
\begin{eqnarray}
	\label{eq:LBQEDaps}
	S_b&=&\int d^4x \bar{\psi}\gamma^{\mu}(\partial_\mu + iA_\mu+ ib_\mu\gamma_5)\psi,
	\end{eqnarray}
can be viewed as layered 2+1-dimensional Dirac theories.
Consider the case when the Weyl node separation is space-like and along the $\hat{z}$ direction. This is equivalent to choosing $b^{\mu}=(0,\mathbf{b})$. Further choosing $\mathbf{b}\parallel \hat{z}$ simplifies our discussion but does not affect the generality of our conclusions. We observe that we can decompose this theory into a sum over two 2+1-dimensional massive Dirac equations. Fourier transforming to momentum space along $\mathbf{b}$ we obtain
\begin{eqnarray}\label{action00}
\nonumber
S&=&\int d^4 x\, \left[\bar{\psi}_L\,\sigma^\mu_{L,\parallel} (\partial_{\mu,\parallel}+i A_{\mu,\parallel})+\sigma_z (k_z+A_z+b_z)) \psi_L\right.\\
&+&\left.\bar{\psi}_R\, (\sigma^\mu_{R,\parallel}\left(\partial_{\mu}
+i A_{\mu,\parallel}\right)-\sigma_z (k_z+A_z-b_z)) \psi_R\right].
\end{eqnarray}
When $A_\mu=0$, the terms $\pm\sigma_z(k_z\pm b_z)$ act as a mass term for 2+1-dimensional Dirac fermions parametrized by $k_z$ with masses $m_{\pm}=(\pm k_z + b_z)$. When $k_z= b_z$ ($k_z= -b_z$), $m_{-} =0$ ($m_{+} =0$) the gap corresponding to chirality $R$ ($L$) closes, setting the location of the 3+1 dimensional $R$ ($L$) Weyl node. 

When $A_\mu\neq0$, a gapped 2+1-dimensional Dirac system with mass $m$ responds with a Hall conductivity $\sigma^{(2D)}_{xy}$ proportional to the sign of its mass, such that $\sigma^{(2D)}_{xy}=\mathrm{sign}(m)e^2/2h$. Depending on the value of $k_z$, the Hall conductivity of the Dirac fermions that compose the Weyl semimetal can either add up or cancel each other (see Fig.~\ref{fig:hall}), resulting in a Hall effect proportional to the Weyl node separation~\cite{Burkov:2011de,Zyuzin2012a,Grushin12,Zyuzin2012,Goswami2013}:
\begin{eqnarray}
\label{eq:dualWeylseffcheck1}
\nonumber
    \sigma^{(3D)}_{xy}&=& \int \dfrac{dk_z}{2\pi} \dfrac{e^2}{h}[\mathrm{sign}(k_z+b_z)-\mathrm{sign}(k_z-b_z)]\\
    &=& \dfrac{e^2}{2h}\dfrac{2b_z}{2\pi}.
\end{eqnarray}
This coincides with the current response derived from Eq.~\eqref{eq:Weyleffb} which we repeat here for convenience (in Minkowski space):
\begin{eqnarray}
\label{eq:dualWeylseffcheck2}
    S_{b} &=& -\dfrac{1}{4\pi^2} \int d^4 x\, \epsilon^{\mu\nu\rho\sigma} b_\mu A_\nu \partial_\rho A_\sigma.
\end{eqnarray}
In Son's 2+1-dimensional duality, a Dirac mass $m$ on one side of the duality maps to $-m$ in the dual theory~\cite{Metlitski,Sachdev:qQdEQuFt}. This means that if our duality is to be correct, we should recover the Hall conductivity resulting from the masses $-m_{\pm} = (\mp k_z - b_z)$. In this case we should recover a Hall conductivity given by
\begin{eqnarray}
\label{eq:dualWeylseffcheck3}
\nonumber
    \sigma^{(3D)}_{\mathrm{dual},xy}&=& \int \dfrac{dk_z}{2\pi} \dfrac{e^2}{h}[-\mathrm{sign}(k_z+b_z)+\mathrm{sign}(k_z-b_z)]\\
    &=& -\dfrac{e^2}{2h}\dfrac{2b_z}{2\pi}.
\end{eqnarray}
In the main text, we showed that for our Weyl duality to hold, $b_\mu$ must map to $-b_\mu$, which is exactly the difference between Eqs.~\eqref{eq:dualWeylseffcheck1} and \eqref{eq:dualWeylseffcheck3}. Hence, our Weyl duality passes this consistency check implied by Son's duality.

\end{document}